# Generative AI in Academic Writing: A Comparison of DeepSeek, Qwen, ChatGPT, Gemini, Llama, Mistral, and Gemma

**Ömer AYDIN, Enis KARAARSLAN, Fatih Safa ERENAY, Nebojša BAČANIN DŽAKULA**

***Abstract -*** *DeepSeek v3, developed in China, was released in December 2024, followed by Alibaba's Qwen 2.5 Max in January 2025 and Qwen3 235B in April 2025. These free and open-source models offer significant potential for academic writing and content creation. This study evaluates their academic writing performance by comparing them with ChatGPT, Gemini, Llama, Mistral, and Gemma. There is a critical gap in the literature concerning how extensively these tools can be utilized and their potential to generate original content in terms of quality, readability, and effectiveness. Using 40 papers on Digital Twin and Healthcare, texts were generated through AI tools based on posed questions and paraphrased abstracts. The generated content was analyzed using plagiarism detection, AI detection, word count comparisons, semantic similarity, and readability assessments. Results indicate that paraphrased abstracts showed higher plagiarism rates, while question-based responses also exceeded acceptable levels. AI detection tools consistently identified all outputs as AI-generated. Word count analysis revealed that all chatbots produced a sufficient volume of content. Semantic similarity tests showed a strong overlap between generated and original texts. However, readability assessments indicated that the texts were insufficient in terms of clarity and accessibility. This study comparatively highlights the potential and limitations of popular and latest large language models for academic writing. While these models generate substantial and semantically accurate content, concerns regarding plagiarism, AI detection, and readability must be addressed for their effective use in scholarly work.*

## 1.    INTRODUCTION

Artificial intelligence has undergone significant development from past to present. In the past, artificial intelligence, which was successful in certain tasks and had limited capabilities, can now provide multitasking or multitalented solutions. These tools, which we can call modern artificial intelligence, have advanced significantly and increased their capabilities in recent years with open source contributions. An important example of these is Google's Transformer architecture in 2017, which laid the technical foundation for later innovations such as OpenAI's GPT series. Similarly, the public availability of BERT on GitHub accelerated its widespread adoption and encouraged the development of tools such as the Transformers library, which democratized access to the latest models. Platforms such as Hugging Face [20] have enabled the sharing of advances in the field. Google even made its 1.6T parameter model accessible on the internet through HuggingFace. With all these developments, the concepts and use of generative artificial intelligence or extended language models have become more widely known and used. This has encouraged not only research with these technologies but also the use of these tools in research. Of course, especially text and image-based production models and ChatBots have begun to be actively used to produce academic content, as they can be used publicly for free or at low cost and open access.

### 1.1 The role AI in Academic Writing

Artificial intelligence (AI) and especially Large Language Models (LLMs) have the potential to revolutionize many sectors. Especially in academic writing, the automatic content generation and information access provided by these models create great opportunities and challenges for both researchers and content producers. Various LLMs are used in text generation, summarization, paraphrasing, writing, language translation and many other areas, and the capabilities of such tools are reshaping the processes of knowledge production and learning.

The issue of AI-generated academic content has led to significant discussions both in academic circles and in industry.  The emergence of models such as ChatGPT, Gemini, Llama, Mistral and most recently Qwen 2.5 Max and DeepSeek v3 has raised the question of how they will affect the production process of academic writing [71, 72, 85, 86]. New-generation artificial intelligence models are trained on larger datasets and can generate higher-quality texts by understanding more complex linguistic structures [81-84]. While popular models such as ChatGPT provide fast and efficient solutions in academic content production, the quality and originality of these solutions have been criticized by many researchers [71-80]. In today's world where the accessibility of these tools has increased, a lot of academic content is produced and published in many academic journals [87-89]. Many journals and publishers create policies on the use of AI-generated content in academic literature or how to use AI tools [90-96]. Detection of this content, depending on the policies, is an important issue. In this context, the evaluation of new models is of critical importance in determining the role of artificial intelligence in academic writing and how these tools can be used in the academic writing process in the future.In our study, a comprehensive comparison will be made between the capabilities of Qwen 2.5 Max, developed by Alibaba, a technology company based in China, and DeepSeek v3, offered by DeepSeek, another Chinese company, in the field of LLM models. Comparing these models with popular AI systems such as ChatGPT, Gemini, Llama, Mistral and Gemma will provide an in-depth analysis of their potential for academic content generation.

### 1.2. Rationale of the study

This study aims to evaluate the academic writing performance of new generation large language models, which have recently been identified with the concept of generative artificial intelligence. We are in a period when their use in academic literature is increasing intensively, and academic journals and editors are experiencing serious difficulties in the face of their incorrect and unethical use. The analysis presented in this study focuses on the evaluation of basic criteria such as the effectiveness of these tools in academic writing tasks, the originality of the content they produce, semantic similarity and readability of the produced texts. In other words, this study investigates the potential usability of AI-generated content in academic writing, tries to provide preliminary information on the detection of misuse and tries to better understand the evolving role of these tools in academic writing. The research extends the methodology used in the studies titled "OpenAI ChatGPT Generated Literature Review: Digital Twin in Healthcare", "Is ChatGPT Leading Generative AI? What is Beyond Expectations?" and "Google Bard Generated Literature Review: Metaverse", which were produced by our research team. It investigates the status of the findings in the above-mentioned studies in current ChatBots while also expanding the scope and results of the study by increasing the evaluation and comparison criteria applied. For example, in one previous study, GPT-3 (via ChatGPT) was used to answer specific questions and restate the abstracts of approximately 40 academic papers. Similarly, in another study, these operations were implemented with Google Bard. In this study, this approach is extended to include a wider set of language models, including more advanced systems such as ChatGPT 4o, Gemini, Qwen3 235B, Qwen 2.5 Max and DeepSeek v3. In addition, instead of comparisons and evaluations made only on plagirisim tool matches, evaluations and comparisons were made with new criteria such as readability, semantic matching and amount of generated content. In this way, the performances of Generative AI tools will be examined in more depth and a higher-level contribution will be created by comparing with previous models.

## 1.3 Research objectives of the study

This study addresses the following key research objectives:

- To evaluate academic writing ability of the Qwen3 235B, Qwen 2.5 Max and DeepSeek v3 by comparing to ChatGPT, Gemini, Llama, Mistral, and Gemma
- To evaluate the extent of originality in academic texts paraphrased by Qwen3 235B, Qwen 2.5 Max and DeepSeek v3, in comparison to other large language models.
- To assess the effectiveness of chatbots in evading detection by AI detection tools.
- To compare the findings of this study with existing research on the academic writing capabilities of AI models, identifying areas of alignment and divergence.

## 1.4 Research Contribution and Novelty

Nowadays, AI tools are used by many researchers in academic literature, consciously, unconsciously, with or without ethical principles. Using these tools is often seen as a normal tool use, and serious confusion has emerged in academic literature about the use of these tools. Since the limits of correct and ethical use have not been fully determined, these tools can be used uncontrolled. At this point, there is a critical gap in the literature regarding the extent to which the use of these tools can be detected and the level of potential to produce content that is original in terms of content, readability and effectiveness. The studies conducted in this field are not sufficient in scope. At the same time, there are very few or very narrow studies on the effectiveness of tools such as DeepSeek and Qwen, which have emerged and have made a great impact in the world. There is also a significant gap in the literature regarding a study comparing these models with existing and popular models. This study contributes to the existing literature by addressing gaps related to the academic writing capabilities of new-generation large language models, especially in terms of content originality, readability, detectability and semantic similarity. This research will comparatively address the performance of new models based on previous studies on academic writing production and will guide on how these models can be used more efficiently in academic content production in the future.

## 2.      THE RISE OF NEW CHINESE ORIGINATED MODELS

Recently developed AI models by China have taken their place among the major language models and have become an important player. They have contributed to the AI ecosystem with the models and technologies they have introduced and continue to do so. Companies such as Alibaba,

DeepSeek and others have taken the lead in developing new generation AI models that challenge models such as OpenAI's GPT series and Google's BERT. These models are increasingly used by more people due to their unique architectures, capabilities and applications and are expanding their application areas.

The rise of Chinese models, especially in the context of natural language processing (NLP), reveals China's increasing influence in the AI field. In this section, we will briefly introduce DeepSeek and Qwen, as they are new and unfamiliar models with their technical features. Both of these models represent the latest innovations in the field of AI, developed by leading Chinese technology companies and gaining significant momentum in various sectors, including academia. Their latest versions, which were announced very recently, are also at a level that can compete with known models with their performance and capabilities. Therefore, we will examine the distinctive features of these models, their technical foundations, and how they compare with other leading AI systems in terms of academic content generation, originality, and readability.

## 2.1. DEEPSEEK Models

DeepSeek emerged as a pivotal player in the development of cost-efficient, large-scale model training. There are significant innovations that enable this efficiency which are described in several papers including [13-16]. We will focus on key models such as DeepSeekMath, DeepSeek-V2, DeepSeek-V3, and DeepSeek-R1, several groundbreaking advancements come to light.

DeepSeekMath model enhanced data quality for superior models. DeepSeekMath demonstrates the importance of high-quality data in model performance. The team developed a FastText-based classifier to filter mathematical content at scale, starting with a robust seed dataset comprising OpenWebMath as positive examples and Common Crawl as negatives. This approach enabled the extraction of 120 billion high-quality math-related tokens, surpassing existing corpora like MathPile and OpenWebMath. The introduction of Group Relative Policy Optimization (GRPO) further optimized training efficiency by eliminating the need for a value model, reducing memory overhead, and accelerating training times. GRPO's cost-effectiveness and scalability were validated through benchmark results, where a 7B model pre-trained on the DeepSeekMath Corpus outperformed larger models, including WizardMath-70B [13].

DeepSeek-V2 model introduced architectural innovations for efficiency. Multi-Head Latent Attention (MLA) is introduced to address the inefficiencies of traditional Multi-Head Attention (MHA) by compressing the Key-Value (KV) cache, reducing memory usage, and speeding up inference without sacrificing accuracy. DeepSeekMoE optimized expert allocation by segmenting experts into specialized units, reducing redundant computations and lowering training costs. Device-limited routing further minimized communication overhead by restricting token interactions to a limited number of devices, enhancing scalability and efficiency [14].

DeepSeek-V3 built on the architectural foundations of previous models, also came with infrastructure breakthroughs. HAI-LLM training framework and the DualPipe algorithm are introduced for pipeline parallelism. DualPipe minimized communication overhead and maximized computation-communication overlap, significantly improving training efficiency. The model also leveraged FP8 mixed precision training to enhance computational speed and reduce memory consumption, while Multi-Token Prediction (MTP) doubled inference efficiency by predicting two tokens in parallel. These innovations enabled DeepSeek-V3 to achieve unparalleled cost-efficiency, outperforming Meta's Llama 3.1 405B in terms of GPU hours, training costs, and energy consumption [15].

DeepSeek-R1 model further reduced training costs by eliminating Supervised Fine-Tuning (SFT) as a preliminary step for Reinforcement Learning (RL). By reintroducing GRPO, the model efficiently learned reasoning tasks without excessive fine-tuning, reinforcing the idea that pre-training provides the core capabilities of a model while fine-tuning optimizes and exposes these capabilities [16]. DeepSeek-R1 represents a significant advancement in AI reasoning models, offering a cost-effective and scalable alternative to traditional training methods. Its rule-based RL framework not only reduces resource requirements but also enhances model alignment with task objectives, paving the way for more efficient and transparent AI development. This underscores the potential of open-source AI to democratize access to high-performance reasoning models while addressing key challenges in reinforcement learning.

DeepSeek's success also depends on its prioritization of meticulous data annotation practices, with organizational leadership, including the CEO, directly engaging in labelling tasks. This emphasis on data quality is further institutionalized in DeepSeek's methodology, as evidenced by the v3 research paper's [15] explicit acknowledgement of annotation contributors, a rarity in technical literature. Such practices underscore the model's alignment with principles of transparency and collaborative development. By openly documenting the scope of its training data while maintaining select proprietary details (e.g., reinforcement learning with human feedback RLHF dataset size), DeepSeek balances open-source accessibility with strategic innovation. These efforts not only advance the technical capabilities of LLMs but also establish benchmarks for ethical data stewardship, challenging prevailing norms in proprietary AI development by emphasizing the foundational role of human expertise in shaping robust, generalizable models. R1 and v3 models exemplify unprecedented transparency in disclosing the scale of human-generated training data for open-source frameworks. The R1 model incorporates a substantial volume of curated datasets, including 600,000 reasoning-specific samples and 200,000 instances of supervised fine-tuning (SFT) data, alongside a human preference dataset for RLHF and synthetic data processed to address cold-start challenges. By addressing the cold-start challenge through structured synthetic data and human expertise, models achieve a functional starting point for training, enabling scalable improvements while mitigating early-stage inefficiencies or biases. This approach underscores the critical role of data quality and human-AI collaboration in overcoming foundational limitations in AI development.

Figure 1 is a scatter plot comparing different AI models based on their MMLU Redux ZeroEval Score (y-axis) and Input API Price per 1M tokens (x-axis) using a logarithmic scale. The chart highlights the performance-to-price ratio, with an optimal range indicated in a shaded area.

DeepSeek-V3 is positioned at the top-left corner, marked with a red star, indicating it has a high-performance score (~89) while maintaining a low input cost. This suggests that DeepSeek-V3 offers one of the best efficiency trade-offs among the models. Other models like Claude 3.5 Sonnet, GPT-4o, and Llama-3.1-405B-Instruct also achieve high performance but may come at a higher cost. Models like DeepSeek-V2.5 and GPT-4o-mini are positioned lower, indicating lower performance scores. The visualization suggests that DeepSeek-V3 has a superior balance of cost-effectiveness and performance compared to its competitors.

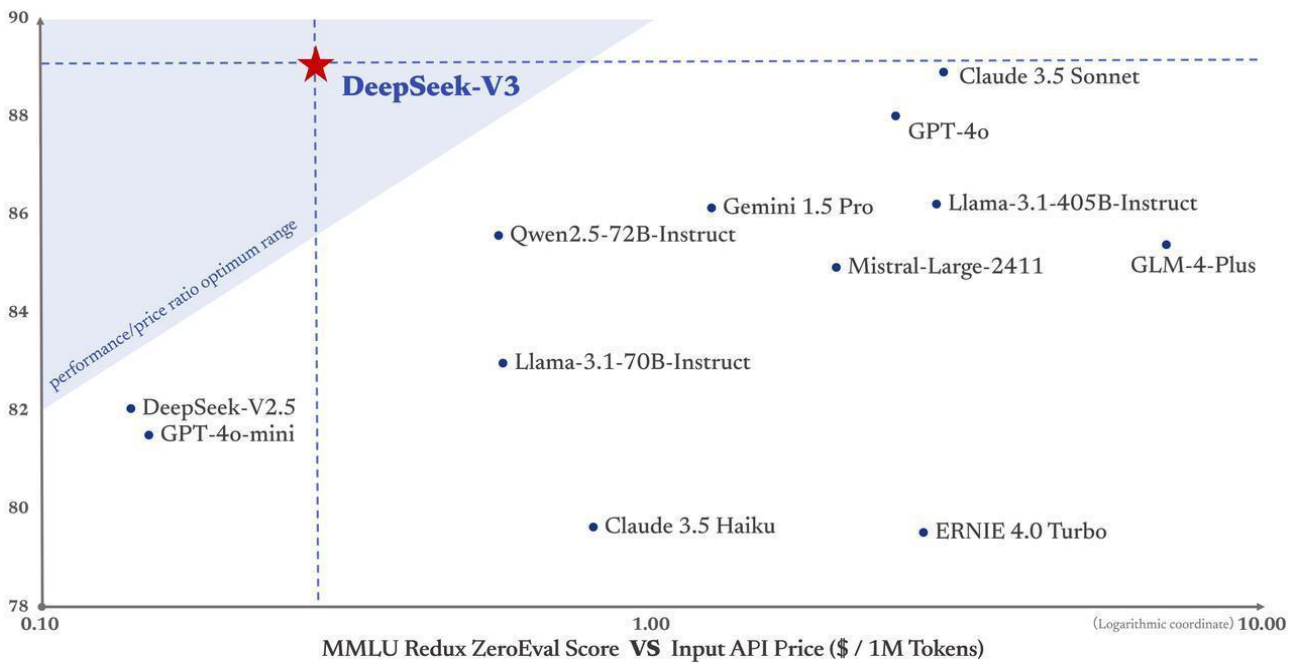

**Figure 1.** Scatter plot comparing different AI models based on their MMLU Redux ZeroEval Score and Input API Price [19]

Figure 2 presents a benchmark comparison table evaluating the performance of multiple AI models, including DeepSeek-V3, Qwen2.5 (72B-Inst.), Llama3.1 (405B-Inst.), Claude-3.5-Sonnet-1022, and GPT-4o-0513. The benchmarks cover various categories such as English language tasks, coding, mathematics, and Chinese language tasks.

DeepSeek-V3 uses a Mixture of Experts (MoE) architecture with 37B activated parameters and a total of 671B parameters, whereas Qwen2.5 and Llama3.1 are dense models. DeepSeek-V3 performs competitively in multiple English benchmarks, scoring highest in MMLU-Redux (89.1), DROP (91.6), and IF-Eval (86.1). In coding, DeepSeek-V3 leads in HumanEval-Mul (82.6) and shows strong performance in Aider-Edit (79.7). DeepSeek-V3 significantly outperforms other models in AIME 2024 (39.2) and CNMO 2024 (43.2) for mathematics. In Chinese tasks, DeepSeek-V3 excels in C-Eval (86.5) and C-SimpleQA (64.1), though Qwen2.5 surpasses it in CLUEWSC (91.4).

DeepSeek-V3 demonstrates strong performance across multiple domains, particularly excelling in English reasoning, coding, and math compared to other AI models in Figure 2.

| **Benchmark (Metric)** | | **DeepSeek-V3** | Qwen2.5 72B-Inst. | Llama3.1 405B-Inst. | Claude-3.5-Sonnet-1022 | GPT-4o 0513 |
|---|---|---|---|---|---|---|
| | Architecture | MoE | Dense | Dense | - | - |
| | # Activated Params | 37B | **72B** | **405B** | - | - |
| | # Total Params | 671B | **72B** | **405B** | - | - |
| | MMLU (EM) | 88.5 | 85.3 | **88.6** | 88.3 | 87.2 |
| | MMLU-Redux (EM) | **89.1** | 85.6 | 86.2 | 88.9 | 88 |
| | MMLU-Pro (EM) | 75.9 | 71.6 | 73.3 | **78** | 72.6 |
| | DROP (3-shot F1) | **91.6** | 76.7 | 88.7 | 88.3 | 83.7 |
| English | IF-Eval (Prompt Strict) | 86.1 | 84.1 | 86 | **86.5** | 84.3 |
| | GPQA-Diamond (Pass@1) | 59.1 | 49 | 51.1 | **65** | 49.9 |
| | SimpleQA (Correct) | 24.9 | 9.1 | 17.1 | 28.4 | **38.2** |
| | FRAMES (Acc.) | 73.3 | 69.8 | 70 | 72.5 | **80.5** |
| | LongBench v2 (Acc.) | **48.7** | 39.4 | 36.1 | 41 | 48.1 |
| | HumanEval-Mul (Pass@1) | **82.6** | 77.3 | 77.2 | 81.7 | 80.5 |
| | LiveCodeBench(Pass@1-COT) | **40.5** | 31.1 | 28.4 | 36.3 | 33.4 |
| | LiveCodeBench (Pass@1) | **37.6** | 28.7 | 30.1 | 32.8 | 34.2 |
| Code | Codeforces (Percentile) | **51.6** | 24.8 | 25.3 | 20.3 | 23.6 |
| | SWE Verified (Resolved) | 42 | 23.8 | 24.5 | **50.8** | 38.8 |
| | Aider-Edit (Acc.) | 79.7 | 65.4 | 63.9 | **84.2** | 72.9 |
| | Aider-Polyglot (Acc.) | **49.6** | 7.6 | 5.8 | 45.3 | 16 |
| | AIME 2024 (Pass@1) | **39.2** | 23.3 | 23.3 | 16 | 9.3 |
| Math | MATH-500 (EM) | **90.2** | 80 | 73.8 | 78.3 | 74.6 |
| | CNMO 2024 (Pass@1) | **43.2** | 15.9 | 6.8 | 13.1 | 10.8 |
| | CLUEWSC (EM) | 90.9 | **91.4** | 84.7 | 85.4 | 87.9 |
| Chinese | C-Eval (EM) | **86.5** | 86.1 | 61.5 | 76.7 | 76 |
| | C-SimpleQA (Correct) | **64.1** | 48.4 | 50.4 | 51.3 | 59.3 |

**Figure 2.** Scatter plot comparing different AI models based on their MMLU Redux ZeroEval Score and Input API Price [19].

## 2.2 QWEN Models

*Qwen* is trained on a large, diverse dataset that covers both general language usage and specialized domains such as science, mathematics, and programming. This carefully curated dataset selection, combined with domain-specific fine-tuning techniques, enhances Qwen's capabilities in areas requiring complex reasoning, such as mathematical proofs and coding tasks [17].

While based on a transformer architecture, Qwen integrates optimizations in the attention mechanism, such as [example if known: multi-query or sparse attention], allowing for more efficient processing of longer text sequences. These enhancements enable the model to achieve comparable or superior performance while reducing computational demands relative to other leading models.

Qwen's training process applies curriculum learning, where the model is initially trained on simpler tasks and progressively exposed to more complex tasks. Mixed precision training is also utilized, significantly reducing training time and improving efficiency. Distributed training and model parallelism further contribute to handling larger datasets effectively [17, 18].

During inference, Qwen benefits from specific optimizations that allow for faster response times, making it well-suited for real-time applications. These strategies collectively make Qwen a resource-efficient and scalable model for both academic and industrial applications.

Figure 3 presents the benchmark comparison across leading models such as Qwen2.5-Max, DeepSeek-V3, Llama-3.1-405B-Inst, GPT-4o-0806, and Claude-3.5-Sonnet-1022. Qwen2.5-Max consistently outperforms other models across Arena-Hard (89.4) and LiveBench (62.2), among other benchmarks.

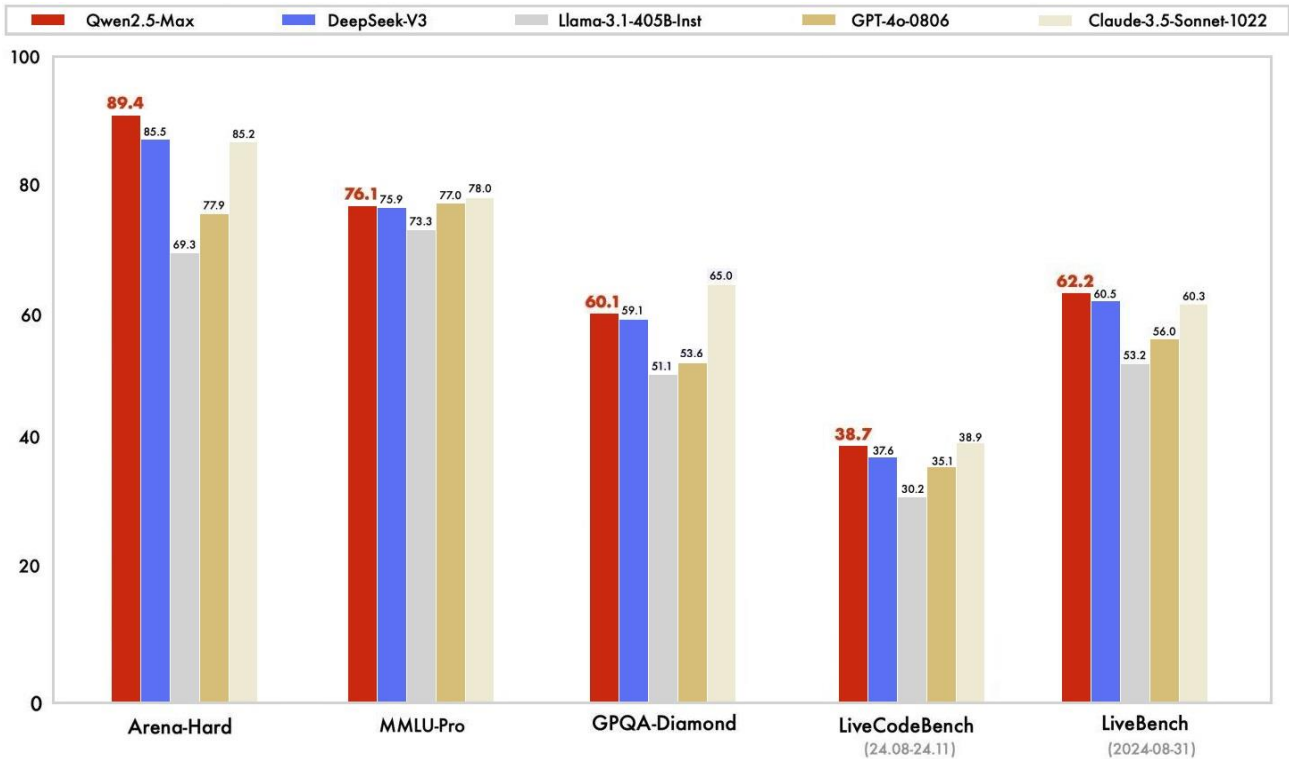

**Figure 3.** Benchmarks of leading models [21]

This consistent leadership highlights Qwen's balanced strength across reasoning, coding, and general knowledge tasks. Notably, DeepSeek-V3 also shows highly competitive results, particularly in reasoning-intensive benchmarks such as Arena-Hard and MMLU-Pro.

On April 29, 2025, a new member joined the Qwen large language model family. It is named Qwen3 and Qwen3-235B-A22B is announced as the flagship. It has competitive results with DeepSeek-R1, o1, o3-mini, Grok-3 and Gemini-2.5-Pro models in coding, math, general abilities, etc. [104]. Competitive performance results can be seen in Figure 4.

| | Qwen3-235B-A22B _MoE_ | Qwen3-32B _Dense_ | OpenAI-o1 _2024-12-17_ | Deepseek-R1 | Grok 3 Beta _Think_ | Gemini2.5-Pro | OpenAI-o3-mini _Medium_ |
|---|---|---|---|---|---|---|---|
| **ArenaHard** | 95.6 | 93.8 | 92.1 | 93.2 | - | 96.4 | 89.0 |
| **AIME'24** | 85.7 | 81.4 | 74.3 | 79.8 | 83.9 | 92.0 | 79.6 |
| **AIME'25** | 81.5 | 72.9 | 79.2 | 70.0 | 77.3 | 86.7 | 74.8 |
| **LiveCodeBench** v5, 2024.10-2025.02 | 70.7 | 65.7 | 63.9 | 64.3 | 70.6 | 70.4 | 66.3 |
| **CodeForces** Elo Rating | 2056 | 1977 | 1891 | 2029 | - | 2001 | 2036 |
| **Aider** Pass@2 | 61.8 | 50.2 | 61.7 | 56.9 | 53.3 | 72.9 | 53.8 |
| **LiveBench** 2024-11-25 | 77.1 | 74.9 | 75.7 | 71.6 | - | 82.4 | 70.0 |
| **BFCL** v3 | 70.8 | 70.3 | 67.8 | 56.9 | - | 62.9 | 64.6 |
| **MultiIF** 8 Languages | 71.9 | 73.0 | 48.8 | 67.7 | - | 77.8 | 48.4 |

1. AIME 24/25: We sample 64 times for each query and report the average of the accuracy. AIME'25 consists of Part I and Part II, with a total of 30 questions.
2. Aider: We didn't activate the think mode of Qwen3 to balance efficiency and effectiveness.
3. BFCL: The Qwen3 models are evaluated in the FC format, while the baseline models are assessed using the highest scores obtained from either the FC or prompt formats.

**Figure 4.** Comparison of performance results of Qwen2 and other models[104]

## 3.  MATERIALS and METHOD

This study adopts a structured methodology to expand upon previous works [31], incorporating a broader range of large language models (LLMs) and more detailed evaluation metrics. The research design involves generating, paraphrasing, and analyzing texts related to the topic of Digital Twin in healthcare, utilizing both cloud-based and locally hosted models. Various tools and criteria were employed to systematically assess originality, readability, AI detectability, and semantic fidelity. The following sections outline the tools, process steps, and evaluation metrics used in the study.

### 3.1. Tools and Environment

For this study, text generation was carried out using the following models: Qwen3 235B, Qwen 2.5 Max, DeepSeek v3, DeepSeek-Coder-v2 16B, ChatGPT 4.0, ChatGPT 4.0 Mini, Gemini 2.5 Pro, Gemini Flash 1.5, Gemma 27B, Llama 3.1 8B, Llama 2 7B, and Mistral 7B.

Local deployment of selected models (DeepSeek-Coder-v2 16B, Llama 3.1 8B, Llama 2 7B, and Mistral 7B) was facilitated using AnythingLLM and Ollama, which handled all related queries and processing tasks. The experiments were conducted on a computer equipped with an AMD Ryzen 7 4800H CPU, 16 GB of RAM, and an Nvidia GeForce GTX 1650i GPU with 4 GB of DDR4 VRAM, ensuring sufficient computational capacity for efficient model operation.

### 3.2. Process Steps

The study will be carried out in the following steps:

Step 1: Text Generation:

- Each model will be used to answer the question "What is Digital Twin?".
- Each model will be asked to create a section text for the topic "Digital Twin in Healthcare". These texts will also be saved separately.
- Each model will paraphrase the abstracts of 40 academic papers (see Table 1).

Step 2: Evaluation:

- Generated texts will be analyzed with iThenticate (plagiarism rates).
- AI detectability will be tested with StealthWriter.ai and Quillbot.com.
- Readability will be checked with Hemingway Editor, Grammarly, and WebFX.

- Semantic similarity will be evaluated with LLMs. Evaluation details can be seen in the following subsection.

## 3.3. Evaluation Metrics

The results of the study will be presented in detailed tables, covering all models. A comparative analysis will be conducted to highlight the differences, strengths, and weaknesses of each model. Furthermore, the performance of ChatGPT from the previous study will be compared with the new results to assess model development over time.

The evaluation of the generated texts focused on the following four main aspects: originality, readability, artificial intelligence detectability, and semantic similarity.

- Originality was assessed using the plagiarism detection tool iThenticate [97]. This tool was selected due to its capability to detect similarities even from content that has been deleted from online sources [98].
- Readability was measured using Hemingway Editor, Grammarly, and WebFX, which evaluate the text's complexity and clarity. These tools were chosen to ensure that the paraphrased and generated texts are understandable, especially when derived from complex scientific abstracts.
- AI Detectability was evaluated through StealthWriter.ai and Quillbot.com. These tools analyze textual features that may indicate whether the content was AI-generated, helping to measure the "human-likeness" of the outputs.
- Semantic Similarity between the original abstracts and the paraphrased outputs was analyzed using ChatGPT 4o, DeepSeek v3, Qwen 2.5 Max and Qwen3 235B. This approach allowed us to assess how well the meaning and semantic structures were preserved.

The abstracts selected for this study (listed in Table 1) were originally published between 2020 and 2022. More recent papers were not evaluated, as older publications are more likely to have detectable matches within plagiarism detection databases. This selection enables a more reliable originality analysis. Additionally, the use of the same papers as in the previous study by Aydın and Karaarslan [31] ensures consistency and enables a comparative analysis of model evolution over time.

**Table 1.** Papers used in the paraphrase process

| # | Authors | Year | Title | Source | Source Type |
|---|---|---|---|---|---|
| [31] | Aydın, Ö., & Karaarslan, E. | 2020 | A Digital Twin-Based Health Information System for The Detection of Covid-19 Symptoms | Online International Conference of COVID-19 (CONCOVID) | Conference |
| [32] | Coorey, G., Figtree, G. A., Fletcher, D. F., & Redfern, J. | 2021 | The health digital twin: advancing precision cardiovascular medicine. | Nature Reviews Cardiology | Journal |
| [33] | Volkov, I., Radchenko, G., & Tchernykh, A. | 2021 | Digital Twins, Internet of Things and Mobile Medicine: A Review of Current Platforms to Support Smart Healthcare. | Programming and Computer Software | Journal |
| [34] | Shengli, W. | 2021 | Is human digital twin possible? | Computer Methods and Programs in Biomedicine Update | Journal |
| [35] | Garg, H. | 2020 | Digital Twin Technology: Revolutionary to improve personalized healthcare | Science Progress and Research | Journal |
| [36] | Popa, E. O., van Hilten, M., Oosterkamp, E., & Bogaardt, M. J. | 2021 | The use of digital twins in healthcare: socio-ethical benefits and socio-ethical risks. | Life sciences, society and policy | Journal |
| [37] | Elayan, H., Aloqaily, M., & Guizani, M. | 2021 | Digital twin for intelligent context-aware IoT healthcare systems. | IEEE Internet of Things Journal | Journal |
| [38] | Gupta, D., Kayode, O., Bhatt, S., Gupta, M., & Tosun, A. S. | 2021 | Hierarchical federated learning based anomaly detection using digital twins for smart healthcare. | 2021 IEEE 7th International Conference on Collaboration and Internet Computing (CIC) | Conference |
| [39] | Zheng, Y., Lu, R., Guan, Y., Zhang, S., & Shao, J. | 2021 | Towards private similarity query based healthcare monitoring over digital twin cloud platform. | 2021 IEEE/ACM 29th International Symposium on Quality of Service (IWQOS) | Symposium |

## 4. FINDINGS and RESULTS

The length of the produced texts can often be seen as unimportant. When the length of the text to be produced is given to ChatBots, they can produce it accordingly. The number of words and length of the produced text may also be related to the LLM settings. On the other hand, if the produced text is too long, unnecessary outputs in terms of meaning and content and that will affect readability may occur. In this sense, the length of the produced text does not mean anything on its own, but it can be important in terms of readability and semantic similarity with the original text that we will use in our analyses. At this point, it would be useful to examine the lengths of the texts produced by different models with default settings. The data presented in Table 2 includes the number of words and characters in the texts produced on the topics of "*Digital Twin*" and "*Digital Twin in Healthcare*". According to the analysis results, the Qwen 2.5 Max model produced the most words (1222 words) and characters (7371 characters) in total. This was followed by Qwen 3 235B, Gemini 2.5 Pro, DeepSeek v3 and ChatGPT 4o. On the other hand, while the Mistral 7B and Deepseek-coder-v2 16B models produced fewer words and characters, it was seen that Deepseek-coder-v2 16B in particular produced only 174 words and 1133 characters in total. These findings are an important indicator in understanding the scope and level of detail of the content produced by each AI model. Comparing the performances of different models provides an idea about which model may be more suitable, especially for certain usage scenarios.

**Table 2.** Generated Text Counts for Asked Questions

| Model | Question 1: What is Digital Twin? | | Question 2: Create a section text for "Digital Twin in Healthcare" | | Total | |
|---|---|---|---|---|---|---|
| | **Number of Generated Words** | **Number of Generated Characters (No spaces)** | **Number of Generated Words** | **Number of Generated Characters (No spaces)** | **Number of Generated Words** | **Number of Generated Characters (No spaces)** |
| ChatGPT 4o | 252 | 1658 | 441 | 2834 | 693 | 4492 |
| ChatGPT 4o mini | 209 | 1269 | 198 | 1245 | 407 | 2514 |
| Gemini 2.5 Pro | 436 | 2406 | 576 | 2782 | 1012 | 5188 |
| Gemini 1.5 Flash | 229 | 1479 | 340 | 2207 | 569 | 3686 |

| Qwen 3 235B | 461 | 2916 | 484 | 3095 | 945 | 6011 |
| Qwen 2.5 Max | 659 | 3832 | 563 | 3539 | 1222 | 7371 |
| DeepSeek v3 | 204 | 1290 | 539 | 3416 | 743 | 4706 |
| Deepseek-coder-v2 16B | 40 | 253 | 134 | 880 | 174 | 1133 |
| Llama 3.1 8B | 124 | 682 | 401 | 2553 | 525 | 3235 |
| Llama 2 7B | 158 | 930 | 256 | 1653 | 414 | 2583 |
| Gemma 27B | 346 | 2115 | 239 | 1576 | 585 | 3691 |
| Mistral 7B | 60 | 331 | 343 | 2053 | 403 | 2384 |

We analyze the number of words and characters in the texts produced by different AI models to paraphrase the abstract sections of the 40 papers we selected. According to the data presented in Table 3, it is seen that the original document contains a total of 7341 words and 42,326 characters. Differences were detected in the paraphrased texts produced by the AI models. While the Qwen 3 235B model produced the most words (7037 words) and characters (43648 characters), Qwen 2.5 Max, ChatGPT 4o mini and DeepSeek v3 exhibited close performances. On the other hand, the Llama 3.1 8B model had the lowest word count, producing only 2615 words and 16,178 characters. These findings reveal that each AI model follows different approaches when paraphrasing the original texts and to what extent they maintain the content density in this process. In particular, it was observed that some models produced shorter and more concise expressions, while others produced more detailed and comprehensive outputs.

**Table 3.** Generated Text Counts for Paraphrased abstract

| Model | Number of Generated Words | Number of Generated Characters (No spaces) |
|---|---|---|
| Original Document | 7341 | 42326 |
| ChatGPT 4o | 6241 | 39657 |
| ChatGPT 4o mini | 6455 | 39538 |
| Gemini 2.5 Pro | 4334 | 28558 |
| Gemini 1.5 Flash | 5950 | 36605 |
| Qwen 3 235B | 7037 | 43648 |
| Qwen 2.5 Max | 6589 | 41397 |
| DeepSeek v3 | 6446 | 40371 |
| Deepseek-coder-v2 16B | 5955 | 37274 |
| Llama 3.1 8B | 2615 | 16178 |
| Llama 2 7B | 3311 | 19837 |
| Gemma 27B | 4937 | 31229 |
| Mistral 7B | 4182 | 25432 |

We analyze the plagiarism rates obtained through iThenticate for texts created by paraphrasing paper abstracts and the answers given by models of different major language models to the questions. According to the data presented in Table 4, the plagiarism rates obtained in question-answer processes vary between 1% and 39%, while these rates in paraphrased abstracts vary between 9% and 57%. In particular, the plagiarism rate in texts paraphrased by ChatGPT 4o mini was determined as 57%, which is the highest rate among all models. In contrast, the plagiarism rate in texts paraphrased by Llama 3.1 8B model remained at the lowest level at 9%. In addition, it is observed that the Gemini 2.5 Pro model has 1%, Qwen 3 235B 7% plagiarism rate for the generated answer texts for questions. On the other hand, Deepseek-coder-v2 16B model exhibits relatively low plagiarism rates in both question-answer (19%) and paraphrase (38%) processes. These findings show that there are significant differences in the originality levels of the content produced by different models, and provide insight into which models are more reliable, especially in paraphrasing processes. These changes in plagiarism rates reveal that each model's rephrasing abilities and fidelity to the source material differ.

**Table 4.** Generated Texts Plagiarism Results

| Model | Text | iThenticate Plagiarism Check Rate (Matching Rate) |
|---|---|---|
| ChatGPT 4o | Question & Answer | 26% |
| | Abstract Paraphrase | 46% |
| ChatGPT 4o mini | Question & Answer | 30% |
| | Abstract Paraphrase | 57% |
| Gemini 2.5 Pro | Question & Answer | 1% |
|  | Abstract Paraphrase | 21% |
| Gemini 1.5 Flash | Question & Answer | 39% |
|  | Abstract Paraphrase | 17% |
| Qwen 3 235B | Question & Answer | 7% |
|  | Abstract Paraphrase | 25% |
| Qwen 2.5 Max | Question & Answer | 29% |
|  | Abstract Paraphrase | 47% |
| DeepSeek v3 | Question & Answer | 37% |
|  | Abstract Paraphrase | 47% |
| Deepseek-coder-v2 16B | Question & Answer | 19% |
|  | Abstract Paraphrase | 38% |
| Llama 3.1 8B | Question & Answer | 36% |
|  | Abstract Paraphrase | 9% |
| Llama 2 7B | Question & Answer | 30% |
|  | Abstract Paraphrase | 15% |
| Gemma 27B | Question & Answer | 29% |
|  | Abstract Paraphrase | 11% |
| Mistral 7B | Question & Answer | 29% |
|  | Abstract Paraphrase | 45% |

The AI-generated content rates of the paraphrased paper abstracts and the responses produced by LLMs were evaluated using AI detection tools (Quillbot.com and StealthWriter.ai), as shown in Table 5. It was determined that almost all of the texts generated in the question-answer processes were detected as generated by AI (100% or close to it). These rates are different in the paraphrased abstract texts. For example, while the paraphrase outputs of the Llama 3.1 8B and Llama 2 7B models were determined as 64% and 62% AI production on quillbot.com, respectively, stealthwriter.ai reported these rates as 89% and 90%. Similarly, while the paraphrase outputs of the Mistral 7B were 80% on quillbot.com, this rate dropped to 62% on stealthwriter.ai. Despite this, all texts produced by the Models can be detected at a high rate by artificial intelligence detectors.

**Table 5.** AI Detector Results

| Model | Text | quillbot.com (AI Rate) | stealthwriter.ai (AI Rate) |
|---|---|---|---|
| ChatGPT 4o | Question & Answer | 100% | 100% |
|  | Abstract Paraphrase | 96% | 86% |
| ChatGPT 4o mini | Question & Answer | 87% | 95% |
|  | Abstract Paraphrase | 77% | 82% |
| Gemini 2.5 Pro | Question & Answer | 100% | 98.2% |
|  | Abstract Paraphrase | 80.25% | 91.10% |
| Gemini 1.5 Flash | Question & Answer | 100% | 100% |
|  | Abstract Paraphrase | 96% | 97% |
| Qwen 3 235B | Question & Answer | 100% | 98.6% |
|  | Abstract Paraphrase | 54.33% | 94.54% |
| Qwen 2.5 Max | Question & Answer | 96% | 100% |
|  | Abstract Paraphrase | 86% | 93% |
| DeepSeek v3 | Question & Answer | 100% | 100% |
|  | Abstract Paraphrase | 88% | 86% |
| Deepseek-coder-v2 16B | Question & Answer | 100% | 100% |
|  | Abstract Paraphrase | 76% | 74% |
| Llama 3.1 8B | Question & Answer | 100% | 100% |
|  | Abstract Paraphrase | 64% | 89% |
| Llama 2 7B | Question & Answer | 100% | 100% |
|  | Abstract Paraphrase | 62% | 90% |
| Gemma 27B | Question & Answer | 100% | 100% |
|  | Abstract Paraphrase | 100% | 95% |
| Mistral 7B | Question & Answer | 100% | 100% |
|  | Abstract Paraphrase | 80% | 62% |

In Table 6, the readability analysis results of the texts produced by different AI large language models were evaluated using Hemingway Editor, Grammarly, and WebFX tools. Hemingway Editor's readability scores were generally low and evaluated at the "Poor" level for all models. According to Grammarly analyses, sentence length and word complexity vary significantly. For example, the average sentence length in the question-answer output of the Deepseek-coder-v2 16B model was 24.9 words, while this value was measured as 20.6 words for the paraphrase output of Llama 3.1 8B. Grammarly scores should be 60 for good readability. So, the Grammarly scores are given in Table 6 is so low. Moreover, the readability scores provided by WebFX also vary between 3.4% and 25.2%. In particular, the paraphrased abstracts of the Llama 2 7B model has the highest readability rate at 24.8%, while the question-answer output of Deepseek-coder-v2 16B has the lowest rate at 5.8%. WebFX score are over 100 so, estimated scores in the table are also very low.

**Table 6.** Readability Check Results

| Model | Text | Hemingway Editor | | | | | Grammarly | | | | WebFX |
|---|---|---|---|---|---|---|---|---|---|---|---|
| | | X/Total sentences are very hard to read | X/Total sentences are hard to read | # of weakeners | # of words with simpler alternatives | Readability score | Word length | Sentence length | Readability score | Text score | Readability |
| ChatGPT 4o | Question & Answer | 23/44 | 0/44 | 7 | 7 | Poor | 6.5 | 15.2 | 7 | 91% | 12.7% |
| | Abstract Paraphrase | 233/289 | 25/289 | 95 | 47 | Poor | 6.2 | 21.3 | 10 | 91% | 13.1% |
| ChatGPT 4o mini | Question & Answer | 14/20 | 3/20 | 6 | 5 | Poor | 6 | 20.1 | 15 | 92% | 23.5% |
| | Abstract Paraphrase | 223/290 | 31/290 | 128 | 48 | Poor | 6 | 22.4 | 16 | 86% | 20% |
| Gemini 2.5 Pro | Question & Answer | 29/64 | 5/64 | 15 | 5 | Poor | 6 | 14.7 | 20 | 95% | 25.2% |
| | Abstract Paraphrase | 153/163 | 5/163 | 64 | 30 | Poor | 6.4 | 26.6 | -2 | 94 | 3.4% |
| Gemini 1.5 Flash | Question & Answer | 16/43 | 0/43 | 10 | 3 | Poor | 6.4 | 12.9 | 12 | 93% | 12.9% |
| | Abstract Paraphrase | 174/387 | 36/387 | 91 | 52 | Poor | 6.1 | 14.9 | 19 | 92% | 22.2% |
| Qwen 3 235B | Question & Answer | 36/87 | 7/87 | 8 | 5 | Poor | 6.2 | 12.9 | 10 | 67% | 22.4% |
| | Abstract Paraphrase | 248/295 | 18/295 | 113 | 52 | Poor | 6.1 | 23.9 | 10 | 82% | 12.4% |
| Qwen 2.5 Max | Question & Answer | 38/78 | 4/78 | 18 | 11 | Poor | 6 | 15.1 | 18 | 87% | 23.2% |
| | Abstract Paraphrase | 240/303 | 26/303 | 110 | 56 | Poor | 6.1 | 21.8 | 12 | 89% | 15% |
| DeepSeek v3 | Question & Answer | 23/53 | 3/53 | 10 | 3 | Poor | 6.3 | 13.6 | 15 | 88% | 20.5% |
| | Abstract Paraphrase | 231/289 | 23/289 | 116 | 54 | Poor | 6.1 | 22.4 | 11 | 87% | 14.6% |
| Deepseek-coder-v2 16B | Question & Answer | 6/7 | 0/7 | 4 | 1 | Poor | 6.4 | 24.9 | 3 | 90% | 5.8% |
| | Abstract Paraphrase | 212/241 | 14/241 | 107 | 54 | Poor | 6.1 | 24.7 | 9 | 84% | 12.7% |
| Llama 3.1 8B | Question & Answer | 17/29 | 3/29 | 6 | 5 | Poor | 6.2 | 17.5 | 15 | 89% | 10.8% |
| | Abstract Paraphrase | 95/127 | 16/127 | 37 | 19 | Poor | 6 | 20.6 | 17 | 88% | 21.6% |
| Llama 2 7B | Question & Answer | 13/23 | 3/23 | 4 | 4 | Poor | 6.2 | 17.6 | 17 | 91% | 11% |
| | Abstract Paraphrase | 126/153 | 16/153 | 62 | 17 | Poor | 5.8 | 21.6 | 22 | 87% | 24.8% |
| Gemma 27B | Question & Answer | 20/39 | 4/39 | 11 | 7 | Poor | 6.1 | 14.9 | 19 | 94% | 15.7% |
| | Abstract Paraphrase | 200/254 | 19/254 | 73 | 30 | Poor | 6.2 | 19.4 | 15 | 94% | 17.2% |
| Mistral 7B | Question & Answer | 14/18 | 4/18 | 6 | 0 | Poor | 5.8 | 22.4 | 22 | 87% | 22.4% |
| | Abstract Paraphrase | 152/179 | 9/179 | 73 | 31 | Poor | 5.9 | 23.4 | 16 | 83% | 19.4% |

According to the data presented in Table 7, it is seen that the paraphrase outputs of all models exhibit a high level of semantic similarity with the original texts. Paraphrased texts produced by models such as Qwen 3 235B, ChatGPT 4o, ChatGPT 4o mini, DeepSeek v3, Gemini 1.5 Flash,

Gemma 27B, Llama 2 7B, Llama 3.1 8B, Mistral 7B and Qwen 2.5 Max generally received a semantic similarity score above 90% in the evaluations made with ChatGPT, DeepSeek v3, Qwen 2.5 Max and Qwen 3 235B models used as semantic checker tools.

Texts paraphrased by ChatGPT 4o have a semantic similarity rate of 89% with ChatGPT, 95% with DeepSeek v3, 98% with Qwen 2.5 Max and 96% with Qwen 3 235B. Similarly, the paraphrase outputs of the Mistral 7B model showed 96.12% semantic similarity with ChatGPT, 85% with DeepSeek v3, 94% with Qwen 2.5 Max and 90% with Qwen3 235B. As can be understood from this information, it has been determined that similar but different semantic similarity rates were calculated in the evaluations made with different tools for the same text. The lowest similarity rate was found when the Mistral model's output was evaluated with the DeepSeek v3 tool (85%). However, even this rate shows that the semantic integrity is largely preserved.

**Table 7.** Semantic similarity Score

| First Content Text | Compared Text | Semantic Similarity Tool Results | | | |
|---|---|---|---|---|---|
| | | chatGPT | DeepsSeek v3 | Qwen 2.5 Max | Qwen 3 235B |
| Papers' original abstract texts | Paraphrased Abstract texts by chatGPT 4o | 89% | 95% | 98% | 96% |
| | Paraphrased Abstract texts by chatGPT 4o mini | 91% | 95% | 96% | 98% |
| | Paraphrased Abstract texts by DeepSeek v3 | 92% | 98% | 97% | 95% |
| | Paraphrased Abstract texts by DeepSeek v2 16B | 91% | 97% | 95% | 93% |
| | Paraphrased Abstract texts by Gemini 2.5 Pro | 91.2% | 88% | 95% | 97% |
| | Paraphrased Abstract texts by Gemini 1.5 Flash | 93.47% | 96% | 94% | 96% |
| | Paraphrased Abstract texts by Gemma 27B | 92.06% | 98% | 96% | 94% |
| | Paraphrased Abstract texts by Llama 2 7B | 90.21% | 97% | 96% | 91% |
| | Paraphrased Abstract texts by Llama 3.1 8B | 86.88% | 98% | 95% | 92% |
| | Paraphrased Abstract texts by Mistral 7B | 96.12% | 85% | 94% | 90% |
| | Paraphrased Abstract texts by Qwen 3 235B | 94.2% | 92% | 98% | 93% |
| | Paraphrased Abstract texts by Qwen 2.5 Max | 96.56% | 90% | 97% | 91% |

## 5.    DISCUSSION and CONCLUSION

Beyond facilitating collaborative progress, open-source practices serve as a critical safeguard against the centralization of AI capabilities within a limited corporate sphere. Decentralizing access to foundational models mitigates risks associated with monopolistic control, such as the imposition of proprietary biases or unilateral governance over transformative technologies. By ensuring equitable access to shared tools, open-source frameworks promote a more inclusive ecosystem where diverse stakeholders can contribute to and scrutinize AI development, thereby fostering transparency and reducing reliance on centralized entities. This approach not only advances technical frontiers but also aligns with ethical imperatives to distribute AI's societal benefits and risks equitably. DeepSeek's performance lies in its systematic approach to efficiency, combining smarter data extraction, optimized architectures, advanced training techniques, and the elimination of unnecessary computations. These innovations not only reduce costs but also set new standards for scalable and cost-effective AI training. As the AI field continues to evolve, DeepSeek's breakthroughs raise important questions about the future of model scaling and the potential for smaller entities to compete with industry giants.

This study provides a detailed evaluation of various generative AI models, focusing on their academic writing capabilities. The discussion focuses on the key findings in the results section, including word counts produced, plagiarism detection, AI detectability, readability, and semantic similarity metrics. Our findings regarding our research questions provide important clues about the role these models can play in different usage scenarios.

Qwen 2.5 Max and DeepSeek v3 provided comprehensive and detailed outputs in the texts they produced, especially on the topics of "*Digital Twin*" and "*Digital Twin in Healthcare*". As seen in Table 2, Qwen 2.5 Max produced the most words (1222 words) and characters (7371 characters) in total, while Qwen 3 235B showed similar performance. This shows that these models are more effective especially on knowledge-intensive tasks. However, some models such as Mistral 7B and Deepseek-coder-v2 16B produced more concise expressions, suggesting that they may be more suitable in certain scenarios.

Although the rates vary, the plagiarism rate allowed in some institutions and universities is set at 10-15% [99]. Some journals allow match rates up to 20% [100]. In most cases, the analysis of the matches in the report is also done along with the rate. Block matches or matches with a single source can be considered [101]. According to the plagiarism analysis results (Table 4), ChatGPT 4o mini has the highest plagiarism rate (57%) and both Qwen 2.5 Max and DeepSeek v3 exhibited moderate plagiarism rates (47% and 47%) in paraphrasing processes, while Llama 3.1 8B model had the lowest plagiarism rate (9%) in abstract paraphrasing. Also, for text generated as answer to the asked questions have usually high plagiarism rates but Gemini 2.5 Pro(1%) and Qwen3 235B (%7) have acceptable rates for academic world. These findings reveal that some models differ in their paraphrasing abilities and their dependence on the source material. In particular, the low plagiarism rates of Gemini 2.5 Pro, Qwen 3 235B, Llama 3.1 8B suggest that these models may be more reliable in paraphrasing processes.

According to the analysis results of AI detector tools (Table 5), it was determined that almost all of the texts produced in question-answer processes (very high) were created by artificial intelligence. Although there are differences in the two measurement tools used, quillbot.com and stealthwriter.ai, the results are generally consistent. For example, while the paraphrase outputs of the Llama 3.1 8B and Llama 2 7B models were determined as 64% and 62% artificial intelligence production on quillbot.com, respectively, stealthwriter.ai reported these rates as 89% and 90%. From here we understand that different tools measure with different methods. However, when we examine the results we obtain comparatively, some models leave less traces of artificial intelligence and therefore can produce more natural or human-like texts.

In this research the readability rates are also evaluated. Grammarly, Hemingway Editor and WebFX are used to estimate the readily of the generated texts. It should be known that Grammarly readability scores should be near 60 for well-readable texts [102] and WebFX readability scores are over 100 [103] to analyse the readability scores in this study. Also, Hemingway Editor gives a detailed score and a result. According to this information, all models tend to use language that is generally complex and difficult to read. According to the Hemingway Editor results, it was determined that sentences both in question-answer and paraphrased abstract texts were very difficult to read. While the "Question & Answer" output of the Gemini 2.5 Pro model had the highest rate in the readability score provided by WebFX at 25.2%, the paraphrased abstract output of Gemini 2.5 Pro had the lowest rate at 3.4%. This shows that a model can produce different results in terms of readability in different tasks. The low readability scores should be discussed by linguists and technical adjustments that need to be made for their development should be considered. This may be because LLMs can generally produce long, complex and academic sentences. Readability tests (Flesch-Kincaid, Gunning Fog Index, etc.) tend to reward short and simple sentences with high scores. In addition, Large Language Models use a wide vocabulary and can sometimes include rare words. Overly technical or abstract terms can cause low scores on tests. In addition, LLMs can sometimes use overly detailed and repetitive expressions. While readability tests reward direct and clear expressions, overly explanations can lead to low scores. LLMs sometimes add unnecessary words to make answers more natural and fluent. Readability tests evaluate unnecessary words negatively. Finally, these tests tend to measure readability at the primary or secondary school level. LLMs, on the other hand, can sometimes produce overly analytical, abstract or in-depth answers, reducing readability.

According to semantic similarity analyses (Table 7), the paraphrase outputs of all models remained in strong semantic relationship with the original texts. Although there are results around 80 percent for some models, the general situation is 90 percent and above in the analyses we conducted with 4 different tools. These results show that these models maintain semantic integrity in the information transfer and paraphrase processes and are reliable for this process.

In previous similar study presented by Aydın and Karaarslan [31], plagiarism values of texts written by paper authors and texts generated by ChatGPT were analyzed. In their study, ChatGPT's GPT-3 model was used and the generated texts were evaluated with iThenticate. In light of the findings and results of Aydın and Karaarslan's study, it was determined that the plagiarism tool match rates of the texts written by the authors were low. On the other hand, it was seen that the answers given by ChatGPT GPT-3 to the questions had relatively low match rates. In contrast, the match rates of the paraphrased abstract texts generated by ChatGPT GPT-3 were quite high. If we compare our study with their study, the match rates of our study are relatively lower both in the answers given to the questions and in paraphrased abstract texts. However, these results are still very high compared to the valid/accepted match rates for the academic ecosystem. The matching rates of the paraphrased abstract text and answers to the questions are consistent with the study of Aydın and Karaarslan for ChatGPT 4o and ChatGPT 4o mini. The positive development in the matching rates may be due to the changes and developments in the LLM models after the study conducted by Aydın and Karaarslan more than 2 years ago. When the other models in our study are examined, it can be evaluated that they have generally similar rates, although there are exceptions.

This study reveals that big and latest models such as Qwen 3 235B, Qwen 2.5 Max and DeepSeek v3, Gemini 2.5 Pro generally perform better in academic writing tasks, but each model offers different advantages in different usage scenarios and has disadvantages. In particular, models such as Llama 3.1 8B and Llama 2 7B stand out with their lower plagiarism rates and higher readability scores. In terms of AI detectability, it has been observed that some models can produce more natural texts but they are still detectable. These findings are an important guide in optimizing AI-based text generation processes and deciding which model is more suitable for which scenario.

## LIMITATIONS and FUTURE DIRECTIONS

While this study provides a comprehensive evaluation of various large language models (LLMs) in academic writing tasks, the following limitations are acknowledged:

- Paraphrase tasks were conducted using abstracts from 40 academic articles focusing primarily on Digital Twin technologies in healthcare. While this provides consistency and depth in a specific area, the study could have been conducted with a larger dataset.
- Although the tests were conducted using different tools, and thus the tool dependency was reduced, the tools used in the assessments could be varied. Results could be compared by adding different tools.
- This study used only computer-based tools to assess originality, readability, and semantic similarity. While efficient and scalable, these assessments may not fully capture nuanced human judgments about text quality, coherence, or academic relevance. This is a limitation of the study.
- The study included the latest versions of popular models in the literature. However, the capabilities and behavior of LLMs can evolve rapidly due to ongoing updates and retraining by developers. Therefore, the findings reflect a specific point in time and may not be applicable to future model versions.
- Tools used for AI detection are generally designed to flag distinct patterns and may not be able to detect highly refined AI-generated text. Conversely, they may also produce false positives for well-written human content, making it difficult to draw definitive conclusions about detectability. However, it is possible to draw a general conclusion about very high-rate texts.

This study analyzed the potential effects of large language models (LLMs) on academic writing, revealing in detail the strengths and weaknesses of existing models. The shortcomings and strengths identified by this study can guide future research. Continuing research in this area is critical both to improve the performance of these models and to better understand their impact on academic writing processes. Limitations of the current study are listed above. To address these limitations, Studies should be conducted on how AI-based writing tools can be more effectively integrated with human users. Human-AI collaboration models can improve the quality of academic writing while also considering ethical and legal responsibilities. Study results show that detection rates of AI-generated texts occur at high rates. Future studies can optimize the production processes of models and develop human-like writing styles to reduce these rates. Future developments in more human-like writing styles may make it more common for people to abuse this ability of LLMs. For this reason, studies on ethical and legal rules should be carried out. The impact of LLMs beyond academic writing in other disciplines should also be examined. Such studies will allow us to better understand the broader use cases of models. Further research should be conducted on the ethical and legal dimensions of AI-generated content. Comprehensive studies are needed on issues such as ownership rights, referencing rules, and the potential for fraud in such content.

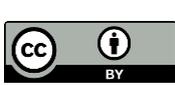